\documentclass[aps,prb,longbibliography,reprint,floatfix,superscriptaddress,nofootinbib]{revtex4-2}

\usepackage{graphicx}
\usepackage{xcolor}
\usepackage[colorlinks=true, urlcolor=blue, citecolor=blue, linkcolor=blue]{hyperref}
\usepackage{dcolumn}
\usepackage{amsfonts}
\usepackage{upgreek}

\begin{document}

\title{Angle-dependent Magnetoresistance of an Ordered Bose Glass of Vortices in YBa$_{2}$Cu$_{3}$O$_{7-\delta}$ Thin Films with a Periodic Pinning~Lattice}
\thanks{The version of record is available free of charge at Condens. Matter , {\bf 8}, 32 (2023); \href{https://doi.org/10.3390/condmat8020032}{https://doi.org/10.3390/condmat8020032}}

\author{Bernd~Aichner}

\author{Lucas Backmeister}
\affiliation{Faculty of Physics, University of Vienna, A-1090, Wien, Austria}

\author{Max Karrer}

\author{Katja Wurster}

\author{Reinhold Kleiner}

\author{Edward Goldobin}

\author{Dieter~Koelle}
\affiliation{Physikalisches Institut, Center for Quantum Science (CQ) and LISA$^+$, Universit\"at T\"ubingen, D-72076 T\"ubingen, Germany}

\author{Wolfgang~Lang}
\email[Corresponding author: ]{wolfgang.lang@univie.ac.at}
\affiliation{Faculty of Physics, University of Vienna, A-1090, Wien, Austria}

\begin{abstract}
The competition between intrinsic disorder in superconducting YBa$_{2}$Cu$_{3}$O$_{7-\delta}$ (YBCO) thin films and an ultradense triangular lattice of cylindrical pinning centers spaced at 30\,nm intervals results in an ordered Bose glass phase of vortices. The samples were created by scanning the focused beam of a helium-ion microscope over the surface of the YBCO thin film to form columns of point defects where superconductivity was locally suppressed. The voltage--current isotherms reveal critical behavior and scale in the vicinity of the second-order glass transition. The latter exhibits a distinct peak in melting temperature ($T_g$) vs. applied magnetic field ($B_a$) at the magnetic commensurability field, along with a sharp rise in the lifetimes of glassy fluctuations. Angle-dependent magnetoresistance measurements in constant-Lorentz-force geometry unveil a strong increase in anisotropy compared to a pristine reference film where the density of vortices matches that of the columnar defects. The pinning is therefore, dominated by the magnetic-field component parallel to the columnar defects, exposing its one-dimensional character. These results support the idea of an ordered Bose glass phase.
\end{abstract}


\maketitle

\section{Introduction}

Magnetic flux penetrates a type-II superconductor in flux quanta of $\Phi_0 = h/(2e)$, where $h$ is Planck's constant and $e$ is the elementary charge. In the ideal situation of a defect-free material, these flux quanta, also known as fluxons or Abrikosov vortices, form the densest packing possible, a hexagonal lattice of flux cylinders. However, in real-world materials, disorder is omnipresent and promotes the nucleation of defects. As a result, they play an essential role in anchoring vortices, and so in preventing their unwanted motion. This motion causes dissipation and spoils the functionality of most superconducting systems.

With the advent of copper-oxide high-temperature superconductors (HTS), there has been renewed interest in the interaction between defect topography and the collective arrangement of vortices. One obvious cause is the strong anisotropy of HTS and their extremely short Ginzburg--Landau coherence lengths. Defects must have dimensions comparable to or greater than the coherence length---smaller ones have a minimal impact. As a result, a short coherence length makes the material more sensitive to microscopic imperfections, such as point defects on the scale of a unit cell or~less.

These novel properties of the HTS give rise to a menagerie of miscellaneous defects with various dimensionalities, hierarchies, correlations, and characteristic length scales. Many different thermodynamic vortex phases can develop when a magnetic field $B_a$ is applied to a superconductor, depending not only on the specific defect topography, but also on the magnitude of the magnetic field, its orientation with respect to the main crystallographic axes, and~the temperature. The very rich phase diagrams of HTS that resulted have stimulated various analyses of this so-called ``vortex matter'' \cite{BLAT94R}.

The fluxons in a clean superconductor are structured regularly with long-range order and three-dimensional (3D) correlation, forming a hexagonal Abrikosov vortex lattice that can be termed a {vortex crystal}. It melts through a first-order transition. Indeed, signatures of a first-order transition have been reported in pure untwinned  YBa$_{2}$Cu$_{3}$O$_{7-\delta}$ (YBCO) single crystals~\cite{SAFA92}.

The inclusion of a sufficiently large density of uncorrelated zero-dimensional (0D) point defects results in a {vortex glass} (VG) \cite{FISH89}. It is akin to a frozen liquid, as the name implies. In such a VG, the vortex cores form bent and even entangled threads hitting as many point defects as possible to minimize their energy. The~VG melts via a second-order transition which can be characterized by critical exponents~\cite{FISH91}. A~key characteristic of a VG is its zero resistance state below the vortex glass melting temperature $T_g$ in the limit of vanishing applied current density $j \rightarrow 0$.

Randomly arranged one-dimensional (1D) pinning centers, such as columnar defects (CDs) formed by swift heavy-ion irradiation~\cite{KRUS94a}, lead to the {Bose glass} (BG), in~which vortex threads are straight and pinned by the columnar defects~\cite{NELS92,NELS93}. However, intrinsic twin boundaries and screw dislocations in YBCO that are oriented parallel to the crystallographic $c$ axis can also cause BG behavior. This has led to inconsistent observations of VG~\cite{WOLT93} or BG~\cite{SAFA96} behavior in YBCO films, depending on the details of the material's morphology. As the VG and the BG have  similar scaling laws for their voltage--current ($V$-$I$) characteristics near the glass temperature, unambiguous discrimination from these measurements alone is impossible. To resolve this issue, transport measurements at different angles of the magnetic field were performed~\cite{GRIG98}, since they indicate the dimensionality of the pinning~centers.

Frustrating disorder in a BG leads to the {vortex Mott insulator} (MI) \cite{NELS93}. Although~the MI also melts through a first-order transition, it differs significantly from the vortex crystal: The latter assembles at any magnetic field in the Shubnikov state between the lower and upper critical magnetic field, whereas the MI is attached to commensurability with a periodic pinning lattice and so emerges only at discrete magnetic fields. Differential resistance measurements in metallic superconductors with periodic pinning potential landscapes have revealed the presence of a MI-to-vortex-liquid transition~\cite{JIAN04,POCC15}.

Finally, in a more realistic experimental setting, the coexistence of an intentionally generated periodic CD lattice and intrinsic defects of various hierarchies might result in the {ordered Bose glass} (OBG) \cite{BACK22}, which is the topic of this research. We will briefly review some of the experimental hallmarks of an OBG before presenting measurements of angular magnetoresistance in nanopatterned YBCO films, which show the predominant pinning along the periodic CDs and support the OBG~picture.

\section{Results and~Discussion}

The competition between the above-mentioned theoretically proposed vortex phases in real materials controlled by temperature, magnetic field, and~disorder has long been an issue~\cite{HWA93,RADZ95,TRAS13}, and~alternative models of vortex melting have been proposed~\cite{REIC00b}. Previous attempts to analyze this problem were hampered by the fact that both columnar and point defects were \emph{randomly} distributed in the material. Only recently has it become feasible to create regular CD patterns with spacings smaller than the London penetration depth using masked~\cite{LANG09,PEDA10,SWIE12,HAAG14} ion irradiation. The discoveries of artificial vortex ice {\cite{TRAS14}} and anomalous metallic states {\cite{YANG19,YANG22}} in YBCO are prominent examples. Even narrower patterns have been reported using focused light-ion irradiation~\cite{AICH19,AICH20,BACK22}.

When the applied magnetic field matches the density of CDs in such densely patterned samples, pronounced vortex commensurability effects can be observed down to low temperatures at so-called matching fields, given by
\begin{equation} \label{eq:match}
    B_m = m\frac{2\Phi_0}{\sqrt{3} a^2},
\end{equation}
where $m$ is a rational number and~$a$ denotes the lattice constant of a triangular CD lattice.  The collective vortex properties are significantly altered at the matching fields $B_m$ \cite{REIC17R}. The most notable commensurability effects are at $m=1$.

\begin{figure*}[t]
\includegraphics[width=\textwidth]{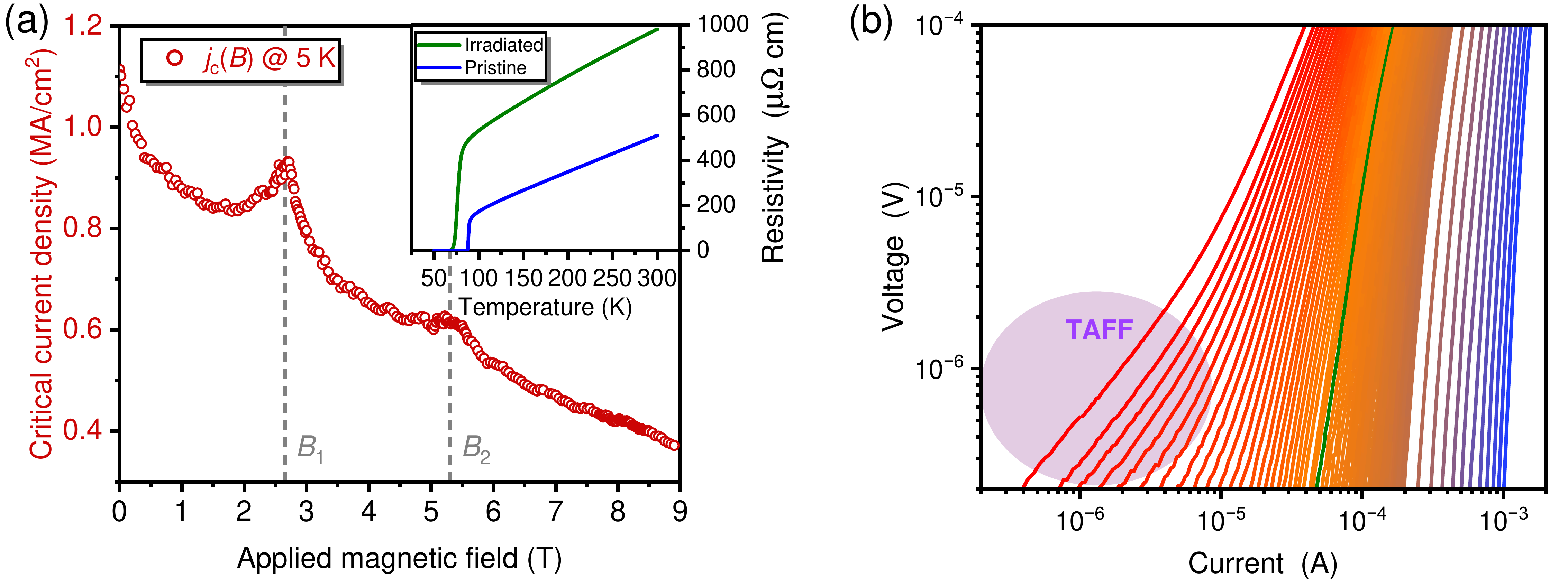}
\caption[]{(\textbf{a}) Critical current density at 5\,K of the irradiated YBCO bridge. The~broken lines represent the number of vortices per unit cell of the triangular CD lattice, which correspond to the matching fields $B_m = m \times 2.65$\,T, as~calculated from Equation~(\ref{eq:match}) with the nominal geometry of the irradiation pattern. Inset: Resistivities of a YBCO thin film with a triangular array of 30 nm spaced columnar defects and an unirradiated reference bridge fabricated on the same substrate.  (\textbf{b}) $V$-$I$ isotherms at 66\,K (red), 65.5 to 52\,K in 0.25\,K steps, and~50 to 30\,K (blue) in 2\,K steps of the nanopatterned sample at the matching field $B_1 = 2.65$\,T. The green line accentuates the isotherm at 59.5\,K, which is closest to the glass temperature $T_g = 59.4$\,K. Figures adapted from~\cite{BACK22}.}
\label{fig:jc}
\end{figure*}

Such a vortex-matching effect is illustrated in Figure~\ref{fig:jc}a in a
 YBCO bridge subjected to focused 30\,keV He$^+$ irradiation in a helium ion microscope. The fabrication process is described in detail elsewhere~\cite{AICH19,BACK22}. The irradiation produces a triangular lattice of CDs with distances $a = (30 \pm 0.6)$\,nm. Superconductivity is completely suppressed inside these CDs, which span the entire thickness of the YBCO~film.

 In perfect agreement with Equation~(\ref{eq:match}), the critical current density $j_c(B)$ as a function of the magnetic field applied orthogonally to the sample surface displays a peak at the first matching field $B_1 = 2.65$\,T and a hump at $B_2 = 5.31$\,T. In~the inset of Figure~\ref{fig:jc}a, the temperature dependence of the resistance in the zero magnetic field of the patterned YBCO film is compared to an unirradiated reference bridge prepared on the same substrate. While the pristine YBCO bridge has a $T_c = 88.7$\,K (defined as the inflection point), irradiation reduces it to $T_c \sim 77$\,K and increases the normal-state resistance.  Some ions scatter away from their direct paths within the planned CDs, resulting in a small number of point defects between the CDs. Such small crystal lattice distortions are known to reduce $T_c$ \cite{WANG95b,SEFR01,LANG10R}. The rising offset of a linear extrapolation of the normal state resistivity, which is substantially larger in the irradiated sample, further supports this. Contrarily, the irradiation has only a marginal effect on the slope, indicating that the charge carrier density in the inter-CD areas is only minimally modified.

The VG and BG theories both anticipate a significant qualitative change in the nonlinear $V$-$I$ isotherms at the magnetic-field-dependent glass-melting temperature  $T_g(B)<T_c$, as shown in Figure~\ref{fig:jc}b for the case of $B_a=B_1$. The green line denotes the power-law behavior of the $V$-$I$ curves near $T_g$ and marks the bifurcation between two sets of the characteristics. On~the left of the green line at $T>T_g(B_1)$, an~ohmic behavior in low currents is connected with the thermally assisted flux motion (TAFF) \cite{KES89} commonly observed in cuprate superconductors.  This suggests that even with vanishing current density $j \rightarrow 0$, the~material has finite resistance. In~sharp contrast, the~isotherms to the right of the green line at $T<T_g(B_1)$ have negative curvature and point to a zero-resistance state at $T<T_g(B_1)$ at $j<j_c$.

Notably, the glass theories anticipate critical scaling of various physical parameters at the continuous second-order phase transition. As~one prominent example, this may be tested by re-plotting the $V$-$I$ isotherms according to
\begin{equation} \label{eq:scalingVG}
(V/I) |1-T/T_g|^{\nu(1-z)} = \mathfrak{F}_{\pm} [(I/T) |1-T/T_g|^{-2\nu}],
\end{equation}
for 3D vortex correlations. In this case, $\nu$ and $z$ are critical parameters that describe the divergence of the glass correlation length and the lifetimes of glassy fluctuations at $T_g$, respectively, \cite{FISH89}. In~a Bose glass, the situation is analogous, and~the respective critical parameters can be easily converted~\cite{WOLT93} to be compatible with Equation~(\ref{eq:scalingVG}).

\begin{figure*}[t]
\includegraphics[width=\textwidth]{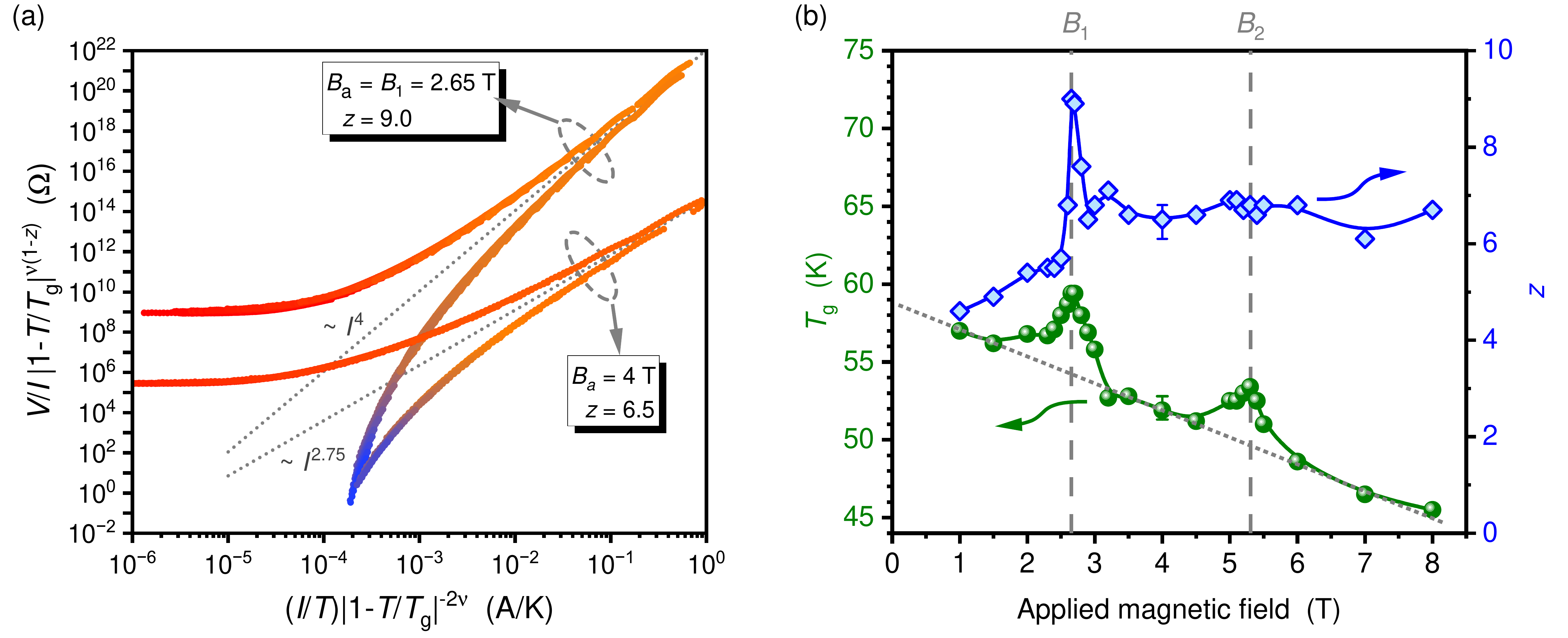}
\caption[]{(\textbf{a}) Comparison of the collapsed $V/I$ vs. $I$ curves using Equation~(\ref{eq:scalingVG}) at the first matching field $B_1 = 2.65$\,T and an off-matching field $B = 4$\,T. The dotted lines indicate the bifurcation and have different exponents. The~ colors are identical to those in Figure~\ref{fig:jc}b. (\textbf{b}) Scaling parameters of the nanopatterned YBCO film. The representative error bars illustrate the uncertainty caused by interdependence of the fit parameters $T_g$ and $z$. Solid lines are guides to the eye. Figures adapted from~\cite{BACK22}.}
\label{fig:compare}
\end{figure*}

The scaling collapse of the $V$-$I$ isotherms onto two material-dependent functions $\mathfrak{F}_{+}$ above and $\mathfrak{F}_{-}$ below $T_g$ has been proven in a number of HTS with varying defect landscapes. The marginal dependence of the critical parameters $\nu$ and $z$ on the applied field $B_a$, however, was a common observation~\cite{LANG96}. The situation is very different in YBCO films patterned with a dense triangular CD lattice, as demonstrated in Figure~\ref{fig:compare}a, where scaling collapses at the matching field $B_1 = 2.65$\,T and at an off-matching field $B = 4$\,T are~compared.

It is instructive to compare the slopes of the bifurcations, represented by dotted lines. The~bifurcation line at $T_g$ not only divides the branches $\mathfrak{F}_{\pm}$ but also obeys a power law
\begin{equation} \label{eq:bifur}
(V/I)\mid_{\scriptscriptstyle T=T_g} \propto I^{(z-1)/2},
\end{equation}
yielding a direct estimate of the dynamic scaling parameter $z$. It is obvious that $z$ and the qualitative behavior of $\mathfrak{F}_{\pm}$ are sensitive to the magnetic field in our nanopatterned samples.

While the critical parameter $\nu = 1.3 \pm 0.2$ is similar in many HTS~\cite{LANG96} and is equally insensitive to $B_a$ here, $T_g$ and $z$ show a strong magnetic-field dependence, as~determined by an extensive set of scaling data at various magnetic fields~\cite{BACK22}. Figure~\ref{fig:compare}b illustrates the decline of the glass temperature $T_g$ with the applied magnetic field, interrupted by maxima at $B_1$ and $B_2$. In contrast, pristine YBCO exhibits a nearly linear reduction of $T_g$ with the magnetic field~\cite{KOCH89}. As long as the applied magnetic field is not commensurable, a similar trend, represented by the dotted line, can be observed in the nanopatterned sample. As~a linear decrease in $T_g$ is theoretically predicted for a BG with random defects~\cite{IKED99c}, the~observed peaks are a distinguishing feature of the~OBG.

The sharp peak of $z$ at the matching field $B_1$ seen in Figure~\ref{fig:compare}b is an even more exciting feature. Remarkably, both VG and BG theories predict, and experiments confirm, $z$ to be roughly in the range $z \sim 4 \dots 6$. The~observed value $z = 9$ definitely suggests a more ordered arrangement of vortex matter, the~{ordered Bose glass}, which is addressed in greater detail elsewhere~\cite{BACK22}. The rather weak rise in $z$ at $B_2$ could be explained by the additional vortices being located at interstitial places, where pinning is dominated by magnetic caging from trapped flux in the CDs and randomly arranged intrinsic pinning sites between the CDs. An~analogous weakening of a BG has been theoretically discussed~\cite{RADZ95}.

The previously discussed results leave one important question unanswered. As the collapse of the $V$-$I$ curves and the resulting parameters cannot reveal the dimensionality of the pinning centers, angular-dependent measurements of resistivity in a magnetic field can provide further information. When the magnetic field is tilted by an angle $\alpha$ off the crystallographic $c$ axis, 0D pinning sites should cause only moderate angle dependence, and 1D columnar defects would effectively pin solely the magnetic field component parallel to them.

According to the scaling approach for anisotropic uniaxial superconductors~\cite{BLAT92}, resistivity curves in magnetic fields applied at an angle $\alpha$ to the $c$ axis collapse into one curve when plotted as a function of a reduced magnetic field:
\begin{equation}
B^* = B_a \sqrt{\gamma^{-2} \sin^2 \alpha + \cos^2 \alpha)},
\label{eq:Blatter}
\end{equation}
where the anisotropy factor $\gamma = \sqrt{m_c/m_{ab}}$, and~$m_{ab}$ and $m_c$ are the effective masses along the $ab$ and $c$ directions, respectively.

\begin{figure*}[t]
\includegraphics[width=\textwidth]{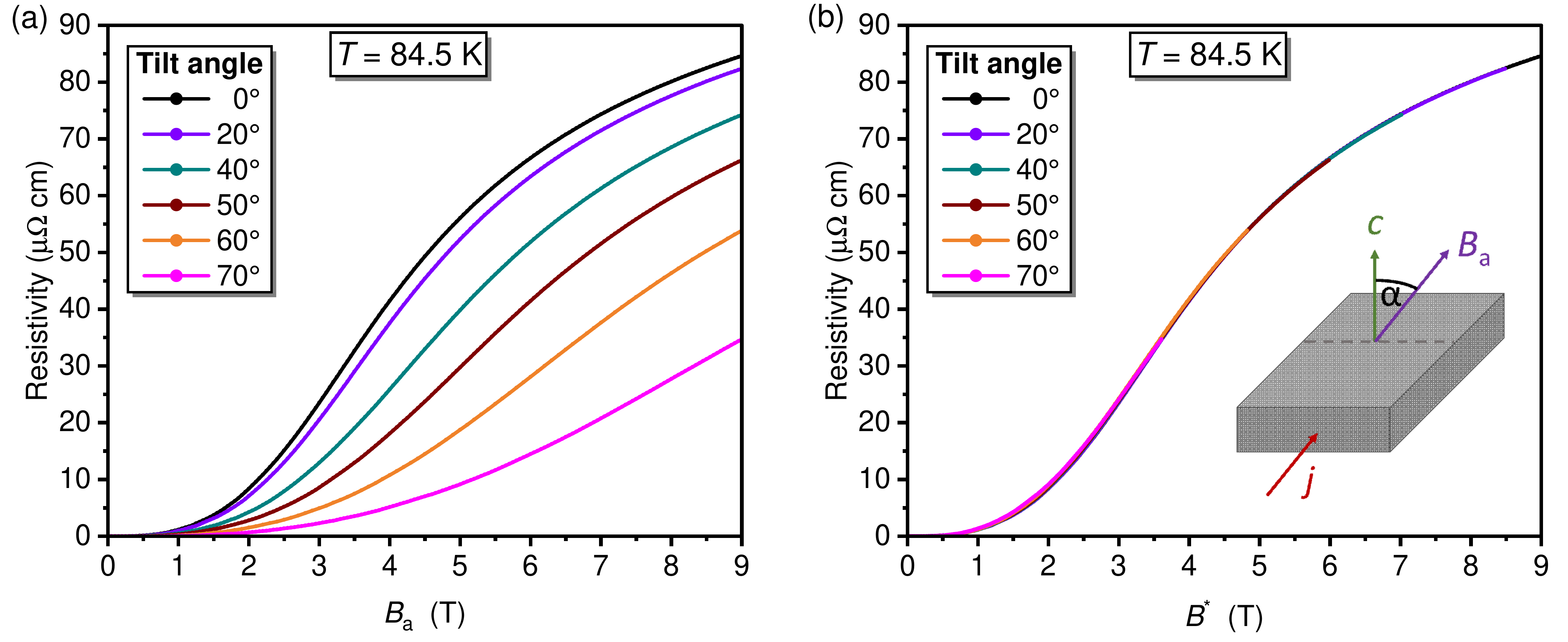}
\caption[]{(\textbf{a}) The angular dependence of the unirradiated reference sample's resistivity as a function of the applied magnetic field, $B_a$. (\textbf{b}) Data scaling to the reduced magnetic field ($B^*$) using Equation~(\ref{eq:Blatter}). The experimental configuration is sketched in the inset.}
\label{fig:ref}
\end{figure*}

Figure~\ref{fig:ref}a shows relevant data on a reference bridge patterned on the same substrate. The~temperature $T = 84.5$\,K corresponds to a reduced temperature $t=T/T_c=0.96$. It was chosen to cover a significant portion of the resistive transition in the applied magnetic field. The~magnetic field's tilt plane was oriented perpendicular to the current direction to achieve a constant Lorentz force condition, as portrayed in the inset of Figure~\ref{fig:ref}b. As~is commonly observed~\cite{IYE90}, the~resistivity at all magnetic fields decreases systematically as $\alpha$ increases. Figure~\ref{fig:ref}b demonstrates the excellent scaling of the data using $\gamma=4.5$ as the only adjustable parameter. In this case, $\gamma$ is slightly lower than estimated from other methods, which could be attributed to the fact that YBCO is a layered superconductor and therefore not perfectly described by an anisotropic model.

\begin{figure*}[ht]
\includegraphics[width=0.6\textwidth]{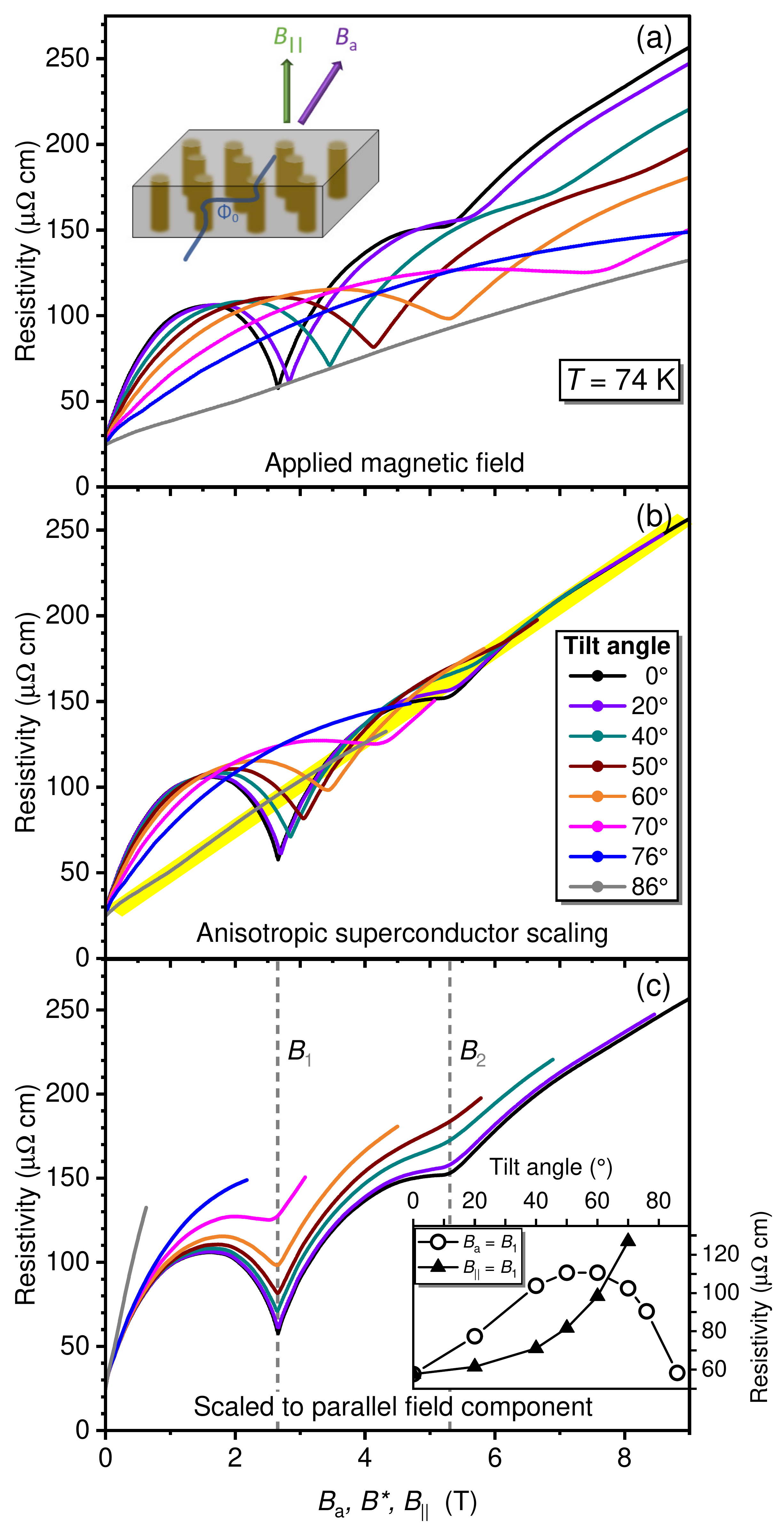}
\caption[]{Angular dependence of the resistivity $\rho$ of the nanopatterned YBCO thin film at 74\,K in various magnetic fields, tilted by an angle $\alpha$. The color code is displayed in the center panel. (\textbf{a})~Resistivity as a function of the applied magnetic field, $B_a$.  The~inset illustrates an example of a kinked flux line (blue) in oblique magnetic fields. Brown cylinders represent columnar defects. (\textbf{b})~$\rho(B^*)$ obtained by scaling according to Equation~(\ref{eq:Blatter}). The~yellow region highlights the collapse of high-field data when $B_a > B_2$. (\textbf{c}) $\rho(B_{||})$ obtained by scaling according to Equation~(\ref{eq:cos}). The~inset shows the angle-dependent resistivity when either the applied field equals the matching field $B_1$ (open circles) or the applied field's parallel-to-CDs component equals $B_1$ (filled triangles).}
\label{fig:angular}
\end{figure*}

Despite numerous studies on the angular dependence of transport properties in superconductors with random 0D and 1D defects, these are rare for periodic CDs in metallic~\cite{METL98,WOMA19} and copper-oxide superconductors~\cite{TRAS13,AICH20}. Figure~\ref{fig:angular}a presents the angular dependence of our nanopatterned sample at $T=74$\,K, which is the same reduced temperature $t=0.96$ as used for the pristine reference sample. In~a magnetic field applied parallel to the CDs and the $c$ axis ($\alpha = 0^\circ$), a~distinct minimum of the resistivity at $B_1$, indicates the commensurability of a plastic flow of vortices~\cite{REIC98} with the CD lattice. A second weaker minimum can be seen at $B_2$. Both attributes correspond to the maxima of $j_c$ shown in Figure~\ref{fig:jc}a.

 When a magnetic field $B_a=B_1$ is applied that is tilted away from the $c$ axis, commensurable locking of fluxons into the CDs is no longer possible, and the paths of magnetic flux lines are distorted. We are not aware of any theoretical work for the specific experimental situation discussed here, but we can draw on findings for randomly arranged CDs~\cite{HWA93,RADZ95}. Several competing effects must be considered in an inclined magnetic field~\cite{HARD96}. (i) The flux lines can be oriented straight along the applied magnetic field; (ii) they can transit through the film along the shortest path (i.e., the $c$ direction), which coincides with being fully locked into the CDs; or~(iii) they can maintain their average directions along $B_a$ and proliferate through the material in vortex segments trapped in the CDs, which are connected by Josephson strings along the CuO$_2$ planes~\cite{HWA93}, as~sketched in the inset of Figure~\ref{fig:angular}a.

 The movement of these Josephson strings is assisted by inherent point defects between the CDs. The Lorentz forces on fluxon segments and Josephson strings point in different directions, causing kinks in the vortex lines to wander, eventually leading to fluxon hopping to neighboring CDs. This process is partially similar to vortex propagation by flux cutting, which has been  observed in vicinal YBCO films~\cite{DURR04}. As a result, as~the number of kinks increases at larger $\alpha$, the resistivity rapidly rises, as~shown by the open circles in the inset of Figure~\ref{fig:angular}c. For~$\alpha > 60^\circ$, a~counteracting reduction in resistivity is presumably caused by intrinsic pinning of Josephson strings between the CuO$_2$ planes~\cite{FEIN90}.

An attempt to scale the data to $B^*$ using Equation~(\ref{eq:Blatter}) fails partially, as~illustrated in Figure~\ref{fig:angular}b. While the high-field data $B_a > B_2$ appear to converge to a universal line (highlighted in yellow), the minima at $B_1$ appear at different values of $B^*$ depending on the tilt angle. Furthermore, $\gamma=2.1$ is much lower than in the pristine YBCO film. The~latter is not surprising, and~it is well-known that nanocomposites in YBCO films significantly reduce anisotropy~\cite{BART19}, even down to $\gamma=1.4$ for a YBCO film with 13\% BaZrO$_3$ (BZO) content~\cite{GUTI07}. At~high $B_a>B_2$, the~scaling of the $\rho(B^*)$ curves appears to indicate that the OBG behavior breaks down and intrinsic point defects dominate the resistivity. In~this instance, the~anisotropic scaling law is~recovered.

The angular-dependent resistivity in samples with a periodic columnar pinning landscape should differ from that in an anisotropic superconductor, as previously found in wider square arrays of CDs~\cite{TRAS13,AICH20}. It can be modeled by formally increasing $\gamma$  to infinity in Equation~(\ref{eq:Blatter}), resulting in
\begin{equation}
B_{||} = B_a \cos \alpha.
\label{eq:cos}
\end{equation}
In this case, only $B_{||}$, the component of $B_a$ parallel to the CD's axes, determines the matching condition of Equation~(\ref{eq:match}). Indeed, when the data are scaled to $B_{||}$, the matching minima in the resistivity are perfectly aligned according to the adapted Equation~(\ref{eq:match}); i.e.,~$B_{||}=B_1$, as~shown in Figure~\ref{fig:angular}c. This finding proves that magnetic flux is trapped predominantly within the CDs even when an oblique magnetic field is applied. As a result of minimizing flux-line kinks, dissipation caused by wandering Josephson strings and fluxon hopping is avoided.

However, this idealized picture must be supplemented by intrinsic defects between the CDs, which, in conjunction with thermal fluctuations, promote vortex line bending. Furthermore, deviations from a path aligned with the applied magnetic field increase the fluxon's energy. At larger angles, these processes become more effective, resulting in enhanced dissipation. The~resistivity minimum at $B_{||}=B_1$ remains visible, but~its base value rises. The~filled triangles in the inset of Figure~\ref{fig:angular}c show that in oblique magnetic fields, resistivity at the matching minima rises only moderately for $\alpha \leq 40^\circ$ but increases progressively at larger~angles.

Remarkably, the~scaled positions of the matching minima remain in place even at large angles $\alpha \leq 70^\circ$. It has previously been established that 2D-glass theories cannot account for the $V$-$I$ data~\cite{BACK22} and that the OBG relies on 3D vortex correlations, which appear to be stable up to high tilt angles.  Indeed, in~2D vortex systems, the~vortex lattice has been found to be fragile in low applied magnetic field~\cite{MACC23}, and~melting was found to be better described by the vortex molasses scenario~\cite{REIC00b}. Note that close to $\alpha = 90^\circ$, additional influences, such as the lock-in of vortices between the CuO$_2$ planes~\cite{TACH89,FEIN90} and possible smectic vortex glass behavior~\cite{RADZ21}, come into~play.

Our results cannot distinguish whether there are two flux quanta inside each CD at $B_2$ or if one is trapped at an interstitial position.  However, the the~scaling of the second minimum suggests that vortices are aligned in parallel to the CDs, even in the latter~case.

The specific scaling of the angle-dependent resistivity in the mixed state of the nanopatterned film affirms the picture of an ordered Bose glass~\cite{BACK22}. The~OBG is a vortex arrangement with order between the vortex Mott insulator and the Bose glass. It is stimulated by a defect landscape composed of both {periodically} arranged 1D columnar defects, capable of trapping one or more vortices along the entire thickness of the sample, and {disordered} defects of diverse dimensionality between the CDs. The latter comprise point defects already present in the pristine material, such as vacancies and intermixing of atom species, and Frenkel defects created by a few ion trajectories scattered off the incident beam direction. They are commonly responsible for VG behavior. On the other hand, 1D screw dislocations, grain boundaries, and~mosaics of 2D twins are examples of intrinsic random defects oriented along the $c$ axis in YBCO that cause behavior.

\section{Materials and~Methods}

The experiments were carried out with very thin YBCO films, epitaxially grown on (LaAlO$_3$)$_{0.3}$(Sr$_2$AlTaO$_6$)$_{0.7}$ (LSAT) substrates by pulsed laser deposition (PLD). Laue oscillations at the YBCO (001) Bragg peak indicated a YBCO film thickness of $t~=~(26.0~\pm~2.4)$\,nm. The~full width at half maximum (FWHM) of $0.08^\circ$ of the YBCO (005) peak's rocking curve confirmed the excellent $c$-axis orientation of the~films.

Following PLD, a~20 nm thick Au film was evaporated in situ with an electron beam. Both the Au and the YBCO films were partially removed using Ar ion milling to form bridge structures  $8\,\upmu$m in width and $40\,\upmu$m in length with voltage probes separated by $20\,\upmu$m. Then, a a~window in the Au layer was removed with Lugol's iodine to allow direct access to the YBCO layer for irradiation while protecting the sample's contact~areas.

The prepatterned YBCO microbridges were introduced into the Zeiss Orion NanoFab He-ion microscope (HIM) and aligned under low ion fluence. The~HIM focused 30\,keV He$^+$ ion beam was set to a spot control value that resulted in an estimated 9\,nm FWHM average diameter for He$^+$ ion trajectories within the film. An~area of $36 \times 16\,\upmu\text{m}^2$ was irradiated with a triangular spot lattice with distances $a = (30 \pm 0.6)$\,nm, covering the entire width of the bridge and extending beyond the voltage probes. A total of $10^4$ ions per spot were required to completely suppress superconductivity in nanopillars that crossed the entire thickness of the YBCO~film.

A too-narrow ion beam can cause unwanted amorphization at the YBCO film's surface, as discussed elsewhere~\cite{AICH19}. In~fact, irradiation of a 26.5\,nm thin La$_{1.84}$Sr$_{0.16}$CuO$_4$ (LSCO) film with a nominally 0.5\,nm wide He-FIB beam revealed orders-of-magnitude-larger damaged areas~\cite{GOZA17}. Contrarily, a comparable experiment in thin-film YBCO bridges resulted in operational Josephson junctions employing an ion fluence, at which no amorphization was detected~\cite{MULL19}. The relative weakness of the copper--oxygen bonds in YBCO as compared to LSCO could explain these differences. Moderate ion doses can suppress the $T_c$ of YBCO while preserving the crystallographic framework. However, in~LSCO, a higher fluence is required to convert the material into an insulator, which causes significantly more damage.

For electronic transport measurements, the contact pads were connected by $50\,\upmu$m thick Au wire and Ag paste to the sample holder of a Physical Properties Measurement System (PPMS) equipped with a 9 T superconducting solenoid and a variable-temperature insert (Quantum Design). At fixed temperatures and in stable magnetic fields parallel to the crystallographic $c$ axis, a large number of $V$-$I$ curves, limited to $100\,\mu$V to avoid heating effects, were collected. A~voltage criterion of 200\,nV was used to define the critical current. A horizontal rotator mounted in the PPMS was utilized to measure angle-dependent resistance. The~$\alpha = 90^\circ$ setting of the dial was calibrated by minimizing the resistance due to the intrinsic vortex lock-in transition when the magnetic field was oriented precisely parallel to the $ab$ planes. All measurements were performed in both current polarities to eliminate spurious thermoelectric signals.

\section{Conclusions}

We studied the interaction of vortices in a landscape of a triangular pinning array of 1D CDs and intrinsic defects in thin YBCO films. Measurements of $V$-$I$ isotherms and resistivity at various temperatures and in oblique magnetic fields revealed a second-order glass transition that we call {ordered Bose glass}. Its characteristics are magnetic-field commensurability effects which are represented by peaks in the glass-melting temperature and the lifetimes of glassy fluctuations. The the~latter exceed theoretical predictions and previous experiments on disordered Bose~glasses.

The frustrated disorder was revealed further when the angular magnetoresistivity was compared to that of a pristine reference sample. Magnetoresistivity scales well in plain YBCO films using the scaling approach for uniaxial anisotropic superconductors. Contrarily, the matching signatures in the nanopatterned sample are determined solely by the magnetic-field component parallel to the 1D pinning channels, indicating that the magnetic flux is trapped within these defect nanopillars. These findings identify the ordered Bose glass as a topological phase intermediate between the vortex Mott insulator and the Bose glass. It differs from the Mott insulator by second-order melting of vortex matter and from the Bose glass by apparent commensurability effects.

The designed periodic pinning landscapes are an excellent test-bed for studying vortex matter in copper-oxide superconductors with their ubiquitous intrinsic defects. They could be useful in experimentally exploring and scrutinizing theoretical predictions of the complex behavior of vortex~matter.

\subsection*{Author contributions}
W.L., D.K., E.G., and~R.K. conceived and supervised the experiments; K.W. grew the film; M.K. patterned the film and performed the focused ion beam irradiation; B.A., L.B., and~W.L. performed the transport measurements; B.A., L.B., and~W.L. analyzed the data; and~all authors discussed the results and contributed to writing the paper. All authors have read and agreed to the published version of the manuscript.

\subsection*{Funding}
This research was funded by a joint project of the Austrian Science Fund (FWF), grant I4865-N; and~the German Research Foundation (DFG), grant KO~1303/16-1. It~is based upon work from COST Actions CA21144 (SuperQuMap), CA19108 (Hi-SCALE), and~CA19140 (FIT4NANO), supported by COST (European Cooperation in Science and Technology).

\subsection*{Data availability}
The data presented in this study are available on reasonable request from the corresponding author.

\subsection*{Conflicts of interest}
The authors declare no conflict of~interest.

\end{document}